# Low-temperature synthesis and electrocatalytic application of large-area PtTe$_2$ thin films

*John B. Mc Manus[1,2], Dominik V. Horvath[1,2], Michelle P. Browne[3], Conor P. Cullen[1,2], Graeme Cunningham[1,2], Toby Hallam[4], Kuanysh Zhussupbekov[5], Daragh Mullarkey[5], Cormac Ó Coileáin[1,2], Igor V. Shvets[5], Martin Pumera[3,6,7], Georg S. Duesberg[1,8], Niall McEvoy\*[1,2]*

[1] School of Chemistry, Trinity College Dublin, Dublin 2, D02 PN40, Ireland

[2] AMBER Centre, CRANN Institute, Trinity College Dublin, Dublin 2, D02 PN40, Ireland

[3] Center for Advanced Functional Nanorobots, Department of Inorganic Chemistry, University of Chemistry and Technology Prague, Technicka 5, 166 28 Prague 6, Czech Republic

[4] Emerging Technologies and Materials Group, School of Engineering, Newcastle University Merz Court, Newcastle Upon Tyne, NE1 7RU. United Kingdom

[5] School of Physics, Trinity College Dublin, Dublin 2, D02 PN40, Ireland

[6] Department of Medical Research, China Medical University Hospital, China Medical University, No. 91 Hsueh-Shih Road, Taichung, Taiwan

[7] Future Energy and Innovation Laboratory, Central European Institute of Technology, Brno University of Technology, Purkyňova 656/123, Brno, CZ-616 00, Czech Republic

[8] Institute of Physics, EIT 2, Faculty of Electrical Engineering and Information Technology, Universität der Bundeswehr, 85579 Neubiberg, Germany

ABSTRACT

The synthesis of transition metal dichalcogenides (TMDs) has been a primary focus for 2D nanomaterial research over the last 10 years, however, only a small fraction of this research has been concentrated on transition metal ditellurides. In particular, nanoscale platinum ditelluride (PtTe$_2$) has rarely been investigated, despite its potential applications in catalysis, photonics and spintronics. Of the reports published, the majority examine mechanically-exfoliated flakes from chemical vapor transport (CVT) grown crystals. While this production method is ideal for fundamental studies, it is very resource intensive therefore rendering this process unsuitable for large scale applications.

In this report, the synthesis of thin films of PtTe$_2$ through the reaction of solid-phase precursor films is described. This offers a production method for large-area, thickness-controlled PtTe$_2$, suitable for a range of applications. These polycrystalline PtTe$_2$ films were grown at temperatures as low as 450 ˚C, significantly below the typical temperatures used in the CVT synthesis methods. To investigate their potential applicability, these films were examined as electrocatalysts for the hydrogen evolution reaction (HER) and oxygen reduction reaction (ORR). The films showed promising catalytic behavior, however, the PtTe$_2$ was found to undergo chemical transformation to a substoichiometric chalcogenide compound under ORR conditions. This study shows while PtTe$_2$ is stable and highly useful for in HER, this property does not apply to ORR, which undergoes fundamentally different mechanism. This study broadens our knowledge on the electrocatalysis of TMDs.

INTRODUCTION

The current interest in nanoscale Group X transition metal dichalcogenides (TMDs) was initially sparked by theoretical papers extolling the promise of PtS$_2$ and PtSe$_2$ for photocatalytic applications.[1-2] PtS$_2$ has proven quite difficult to fabricate while PtSe$_2$ has been grown by a number of methods.[3-5] Similar to other TMDs, the ditelluride of Pt has thus far been less studied than the disulfide or diselenide counterparts. The synthesis of bulk PtTe$_2$ through the reaction of elemental Pt and Te was reported as early as 1897, with a number of later papers examining bulk properties of crystalline PtTe$_2$ grown by chemical vapor transport (CVT).[6-9] However nanoscale, or 2D, PtTe$_2$ is still a relatively unexplored system, with only a few published works in recent years that examine it experimentally.[10-11]

PtTe$_2$ is isostructural with PtSe$_2$ and has a 1T structure, with space group P$\bar{3}$m1 and point group D$_{3d}$. It is semimetallic with notably stronger interlayer forces than many other TMDs, such as MoS$_2$.[12-14] Due to these relatively strong interlayer interactions, the distances between Te atoms in adjacent PtTe$_2$ layers are comparable to those in trigonal and monoclinic tellurium (3.4Å vs 3.5Å), indicating that it would be an oversight to regard the interlayer interactions as purely van der Waals.[13] PtTe$_2$ has been predicted to be a potential type-II Dirac semimetal, similar to type-II Weyl semimetals MoTe$_2$ and WTe$_2$.[15-17] These materials offer interesting systems in which to investigate novel quantum phenomena, and they have been touted for potential applications in terahertz photodetection, spintronics and quantum computing.[18-20] Much of the recent work on PtTe$_2$ has focused on examining it in the context of these exotic electronic properties, with a number of works demonstrating promising results.[21-22] PtTe$_2$ has also been investigated for ultrafast photonic applications.[23]

Similar to other 2D materials, the layered structure of PtTe$_2$ opens up the possibility of high surface area to volume ratio electrodes; as such, it has been examined as an electrocatalyst for a number of processes. PtTe$_2$ was found to have the greatest electrocatalytic performance for the hydrogen evolution reaction (HER) of all the Pt dichalcogenides, primarily due to its high conductivity.[24-25] It has also been reported to be as effective for catalysis of the oxygen reduction reaction (ORR) as Pt/C, while being less toxic in certain circumstances.[26-27] The catalytic activity of the basal plane can be further increased through defect engineering such as oxidation of the surface.[28]

It is evident that the potential of PtTe$_2$ outlined above provides compelling motivation for targeted research into its synthesis on the nanoscale. However, the fabrication of monolayers of PtTe$_2$, through chemical vapor deposition (CVD) or mechanical exfoliation, is challenging due to the aforementioned relatively strong interlayer bonds.[13-14] This also means that making atomically-thin flakes using liquid-phase exfoliation (LPE) is extremely difficult. This is corroborated by the findings of the only paper published thus far on LPE of PtTe$_2$ wherein the flakes produced had lateral dimensions of ~100 nm and average thicknesses of 25 nm.[23]

The majority of experimental studies on PtTe$_2$ published to date have used CVT synthesis methods to produce bulk PtTe$_2$ crystals that can then be exfoliated to make 2D crystals. This is an extremely laborious and time consuming process. CVT synthesis also requires temperatures in excess of 700 ˚C and long dwell times, typically tens of hours.[15, 21, 25, 29-30] These long dwell times at high temperatures make the process industrially unappealing and further processing steps would be required before any potential applications could be considered. $PtTe_2$ has also been synthesized by CVD approaches but uptake of these methods has been hindered by the harsh synthesis conditions. Of the two currently published $PtTe_2$ CVD studies one involved reacting elemental powders at 1150 ˚C, while the other synthesized $PtTe_2$ from $PtCl_4$, NaCl and Te precursors at a reaction temperature of 800 ˚C.[10, 12] Both of these resulted in discrete flakes of $PtTe_2$ with lateral dimensions from hundreds of nanometers to 10 µm and thicknesses above 3 nm. Another method to produce discrete flakes is the eutectic solidification method demonstrated by Hao *et al.*. This involved heating a Pt film and Te powder to 700 ˚C under a forming gas (Ar/$H_2$) atmosphere.[14]

High-quality, thickness-controlled $PtTe_2$ films have been grown on bilayer graphene/6H-SiC (0001) substrates using molecular beam epitaxy (MBE). However, this required ultra-high vacuum conditions along with a precise and complex set of *in situ* anneals to prepare the substrate and the film, limiting its applicability.[31] The exciting potential applications of $PtTe_2$ leave a pressing case for developing a straight-forward synthesis method that yields large-area samples in a scalable manner.

This work examines the synthesis of large-area polycrystalline $PtTe_2$ films through the reaction of pre-deposited films of Pt and Te. The Pt and Te films are deposited by electron-beam evaporation and electrodeposition respectively, yielding large-area films with well-defined geometries. The $PtTe_2$ films were then grown at temperatures as low as 450 ˚C. Characterization indicates the successful synthesis of $PtTe_2$ over a range of conditions with films showing low levels of surface oxidation. Morphological investigations show the tendency of the reactants to dewet from the growth substrate to form thicker crystals due to the relatively strong interlayer interactions in PtTe$_2$. Finally, the electrocatalytic properties of the films are examined for the HER and the ORR with the results indicating that these films are highly catalytically active.

RESULTS AND DISCUSSION

Film Synthesis and Characterization

Polycrystalline PtTe$_2$ films were synthesized by first depositing a Pt metal layer onto a substrate, and then electrodepositing Te on top of this. The substrates were then placed in nested crucibles in a quartz-tube furnace, as shown in the supporting information (SI), Figure S1. The nested crucibles served to maintain a high local concentration of Te in the vicinity of the samples, while minimizing the quantitiy of Te required in the furnace. The formation of PtTe$_2$ took place under a ~700 mbar N$_2$ atmosphere with a dwell time of 90 minutes. The standard synthesis temperature was at 450 ˚C. The effect of the growth temperature is discussed later in the paper. Predeposition of the Pt and Te layers allowed for strong control to be maintained over the reactant quantity on each sample, while also providing intimate contact between the reactants during the reaction. An advantage of this method is that the thickness of the final PtTe$_2$ films can be defined by controlling the thickness of the initial precursor layers. This method also allows large-area, uniform films with defined patterns to be synthesized, which is challenging for CVD-type processes.

PtTe$_2$ films can be synthesized from initial Pt layers of a variety of thicknesses using this method, those between 2 and 40 nm are examined in this work. Films will be referred to by the thickness of the initial Pt film for convenience e.g. a 20 nm PtTe$_2$ film implies a PtTe$_2$ film grown from an initial 20 nm Pt film. The initial growth studies of the PtTe$_2$ were carried out on 20 nm films, with the optimization of the growth temperature and the film thickness being discussed later in the work.

A Raman spectrum of a typical 20 nm PtTe$_2$ film is presented in Figure 1(a). There are two primary peaks visible in the spectrum, centered at ~111 cm$^{-1}$ and ~158 cm$^{-1}$. There are 3 atoms in the unit cell of PtTe$_2$ and its D$_{3d}$ symmetry results in the two Raman-active modes seen here, the in-plane E$_g$ mode at ~111 cm$^{-1}$ and the out-of-plane A$_{1g}$ at ~158 cm$^{-1}$. These are consistent with previously-reported peak positions, and relative intensities, for bulk-like PtTe$_2$.[14-15] This offers initial confirmation of the successful synthesis of PtTe$_2$ using this method.

The Raman spectrum can also give an indication of the quality of the material with the full width at half maximum (FWHM) of Raman modes being linked to the crystallinity of the sample.[32] Both Raman modes were fitted with Lorentzian peaks allowing the FWHM to be extracted, as shown in the SI, Figure S2. Examining a number of films gave an average FWHM for the E$_g$ peak of 6.9 ± 0.5 cm$^{-1}$ and 4.7 ± 0.4 cm$^{-1}$ for the A$_{1g}$ peak. These values are significantly narrower than FWHM of previously reported PtTe$_2$ grown at much higher temperatures, as shown in Table S1 in the SI. This illustrates the samples in this work are of high crystalline quality despite the relatively low synthesis temperature.

X-ray photoelectron spectroscopy (XPS) was used to gain insight into the quality and purity of the PtTe$_2$ films produced. Figure 1(c) and (d) display the spectra of the Pt 4f and Te 3d core-level regions of a 20 nm film of PtTe$_2$. The Pt 4f region shows two sets of doublets, indicating that 90% of the Pt atoms are bound to Te in the form of PtTe$_2$, while 10% are Pt bound to oxygen. Similarly for the Te 3d core levels, 81% are Te bound to Pt as PtTe$_2$, while 19% of the Te atoms are bound to oxygen. While this oxidation level is significant, it is expected to only be present in the first few nanometers of the films as has been shown for previous works on telluride TMDs.[33] This oxidation potentially occurred upon the exposure of the films to atmosphere between synthesis and measurement. Significantly there are no signals corresponding to elemental Pt or Te present in the observed spectra, indicating the complete reaction of the precursors.

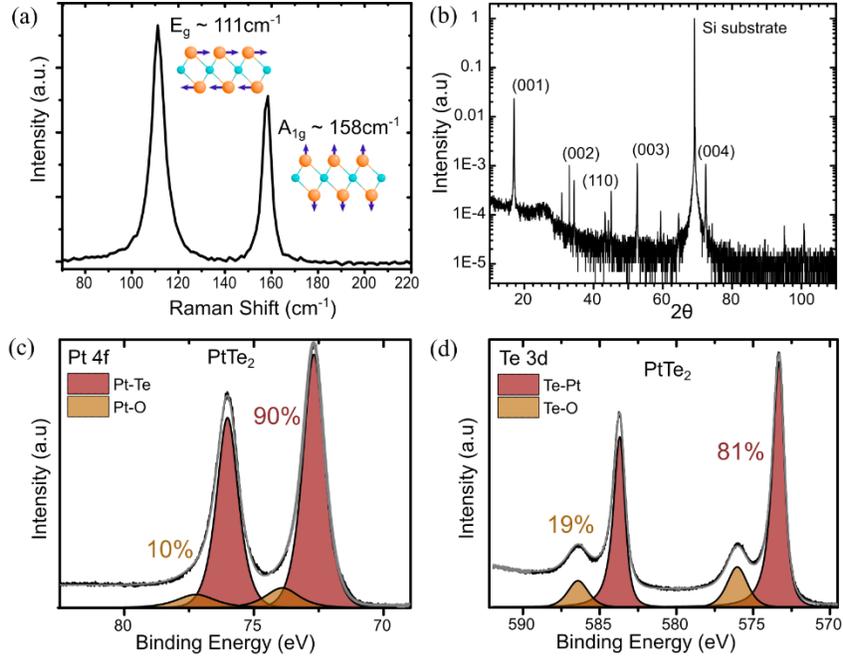

Figure 1(a) Raman spectrum of a 20 nm $PtTe_2$ film showing the two Raman-active vibrational modes, along with ball and stick models representing these. (b) XRD of 20 nm $PtTe_2$ film with the most intense peaks labelled on the graph. (c) & (d) XPS of $PtTe_2$ film showing the Pt 4f and Te 3d core levels respectively. Combined these three techniques serve to confirm the successful synthesis of $PtTe_2$.

The calculated stoichiometry of the $PtTe_2$ on the surface is $PtTe_{1.8}$, signifying a Te deficiency in the lattice. However when all surface oxides are included, the Pt:Te ratio is 1:1.94, much closer to the ideal 1:2 ratio. Finally, the asymmetric shapes of the $PtTe_2$ XPS peaks are indicative of the semimetallic nature of the films.

X-ray diffraction (XRD) allowed the crystallinity of the films to be examined. Figure 1(b) shows the spectrum obtained from a 20 nm $PtTe_2$ film. There are many peaks visible in the spectrum, the most intense of these being associated with the substrate at ~69 °. The other prominent peaks correspond well to the predicted, and previously published, spectrum for $PtTe_2$.[10, 15, 27] The (001), (011), (002), (012), (110), (003), (013), (004) and (121) peaks are all discernible, with the most intense of these being labelled on the graph. In addition to the discrete peaks, there is a significant broad background signal visible in the measurement. This combination of crystalline peaks and a broad background indicates that these are polycrystalline films of PtTe$_2$ with crystalline grains of appreciable size.

A variety microscopy techniques were used to further investigate the morphology of the films. Representative scanning electron microscopy (SEM) and atomic force microscopy (AFM) images of a 20 nm PtTe$_2$ film are shown in Figure 2(a) and (b) respectively. These both show that this synthesis method yields polycrystalline films of PtTe$_2$ with high levels of surface coverage. The dark areas visible in the SEM are the underlying pyrolytic carbon (PyC) substrate. The film has grain sizes of between 200 nm and 1 µm with terraces or steps visible on the surface. This correlates well with what was observed in the XRD.

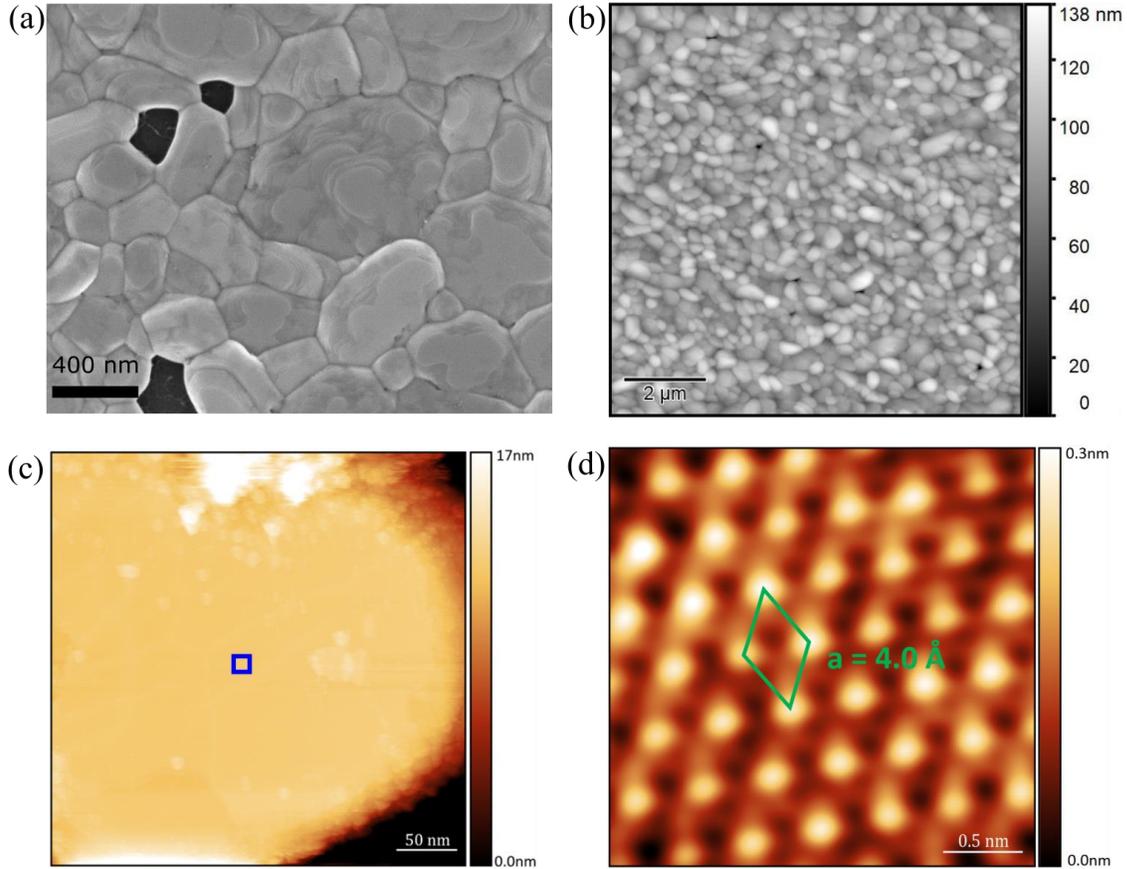

Figure 2 (a) SEM image of 20 nm PtTe$_2$ film. The crystalline grains have visible steps on the surface, while the darker areas are the underlying PyC substrate (b) AFM image of the same sample shown in (a). (c) STM image of a PtTe$_2$ grain indicating some contamination on the surface of the crystal (350×350 nm, V=1.5 V and I=90 pA). (d) Atomic resolution STM image of highlighted area in (c), showing the crystallinity of the sample and the expected interatomic distance of 0.4 nm (2.5×2.5 nm, V=0.7 V and I=300 pA).

AFM analysis, shown in Figure 2(b), displays a polycrystalline surface made up of randomly oriented grains of PtTe$_2$. The extracted RMS roughness of the film is 14.2 nm. Both the SEM and AFM highlight evident 3D qualities of the individual grains, with substantial thicknesses within an order of magnitude of their other dimensions, examined in further detail in Figure S3 in the SI. This is tentatively attributed to the relatively strong interlayer bonding in PtTe$_2$ compared to other layered TMDs, as it was not observed in similarly grown MoTe$_2$ films.[13-14, 34]

Figure 2(c) and (d) show scanning tunneling microscopy (STM) images of an individual grain from a 20 nm PtTe$_2$ film. Figure 2(c) shows a larger-area scan that indicates some contamination present on the surface of the sample, while (d) is an atomic resolution image of the PtTe$_2$ surface, illustrating the crystallinity of the PtTe$_2$ grains. The measured interatomic spacing matches very well to the expected value of 0.4 nm.[14, 35] This combined microscopic analysis allows us to conclude the films are polycrystalline, made up of crystalline grains of PtTe$_2$ up to 1 µm in size.

An advantage of this synthesis method over previously reported CVD and CVT methods is that is gives large-area polycrystalline films of PtTe$_2$. While not suitable for all applications, these large-area films have advantages over samples obtained from other methods in that they have uniform and controllable thickness over defined areas. This makes them more suitable for scaling and integration than the discrete flakes grown by CVD.

Synthesis Optimization

To better understand and optimize this synthesis method, the effect of varying reaction parameters on the resultant PtTe$_2$ films was investigated. First, the dwell temperature of the furnace during the synthesis process was varied. The resulting films were then examined with Raman spectroscopy, shown in Figure 3(a). These samples were grown at temperatures between 350 ˚C and 750 ˚C. A characteristic PtTe$_2$ Raman signal is only seen for samples processed at temperatures between 450 ˚C and 650 ˚C, while at lower temperatures the only visible peaks correspond to Te.[36-37] The 750 ˚C run shows a number of unidentified peaks that are thought to be due to the formation of a less Te-rich phase, this is discussed further in the SI, Figure S4.

Through examination of the synthesis temperature results and the published Pt-Te phase diagram (reproduced in Figure S5 of the SI) the temperature range in which PtTe$_2$ can be successfully fabricated using this method can be rationalized. The melting point of Te is 450 ˚C, this gives an effective lower temperature limit for this process. Below this temperature both the Te and Pt are in the solid phase, thereby restricting effective reaction between the two. The Pt-Te phase diagram indicates that PtTe$_2$ is expected to grow up to 1150 ˚C. However, due to the specifics of the experimental setup used this does not occur. Above 650 ˚C the Te is likely vaporized too quickly to effectively react with the Pt, or any PtTe$_2$ that does form then loses Te to the atmosphere. This hypothesis is supported by previous work on MoTe$_2$ synthesis using a similar method and by the reports of telluride TMDs losing Te at elevated temperature in a Te-deficient environments.[34, 38-40]

This low synthesis temperature of 450 ˚C is an advantage of this method over previous works as it allows the successful growth of PtTe$_2$ with a much lower thermal budget. The majority of experimental studies of PtTe$_2$ have used CVT-grown crystals that typically require temperatures in excess of 700 ˚C,[15, 21, 25, 29-30] while the two CVD studies reported were carried out at temperatures between 800 ˚C and 1150 ˚C.[10, 12] The significantly lower temperature used here opens up the potential for growth on a larger variety of substrates, and integration with more diverse processes.

Another parameter examined was the influence of the relative quantities of Pt and Te deposited on the growth substrate. Shown in Figure 3(b)-(d) are a series of SEM images of PtTe$_2$ films grown from 20 nm Pt precursor films with varying initial quantities of Te. The Te was deposited using a pulsed electrodeposition method, with the films shown having increasing amounts of Te of between 2,500 and 10,000 pulses. The thickness of Te these equate to is discussed in the SI, Figure S6. From Figure 3 (b)-(d) it is clear that the morphology of the resultant PtTe$_2$ films is influenced by the initial quantity of Te present on the sample. Films with a lower Te:Pt ratio have high levels of coverage of PtTe$_2$ crystals visible on the surface but have smaller grain sizes. Films with more Te initially have larger grains but the crystals do not cover the entire surface. This is discussed further in the SI, Figure S6.

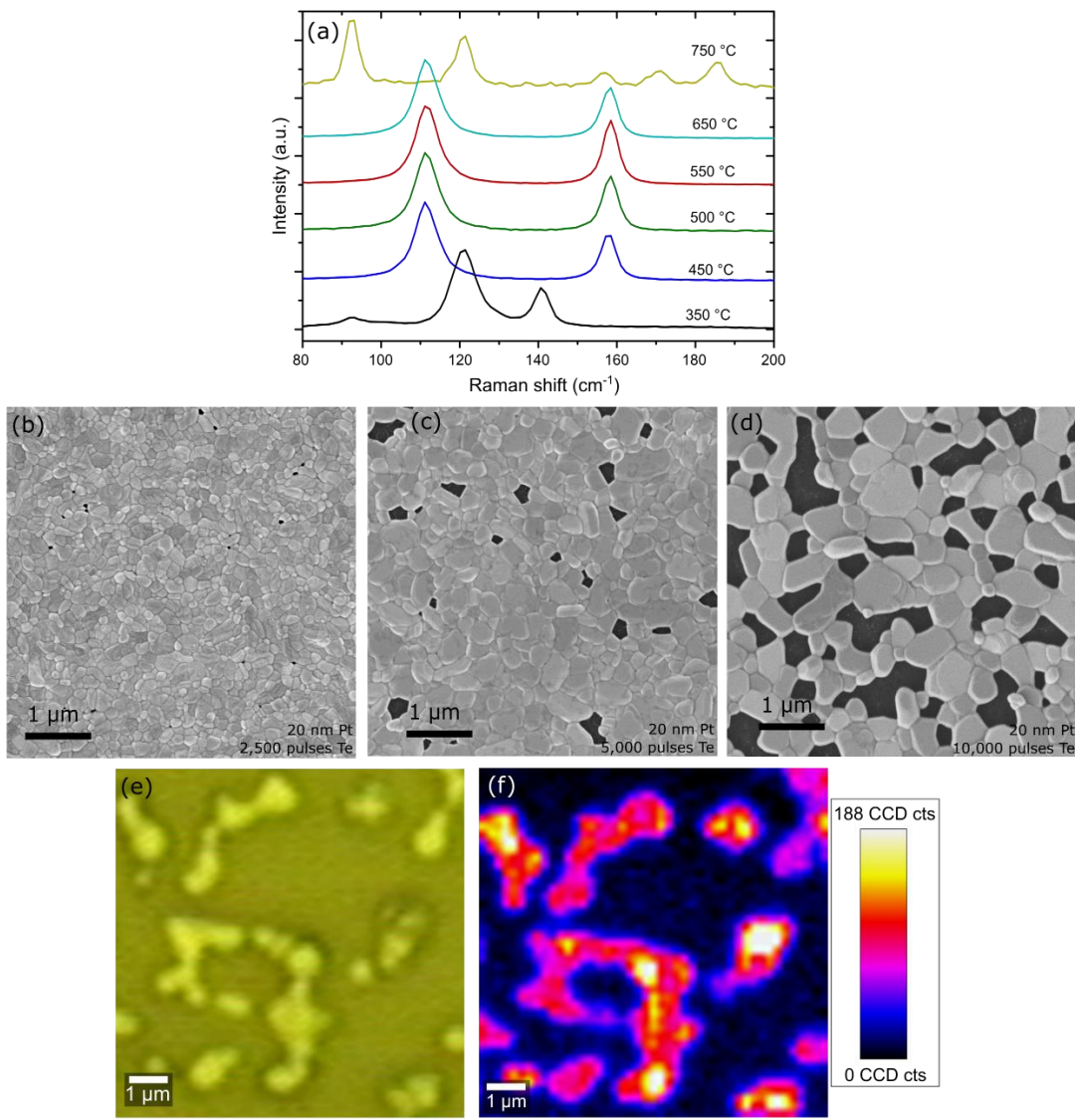

Figure 3(a) Raman spectra of films grown at different temperatures indicating the range in which PtTe$_2$ can be synthesized using this method. (b) – (d) SEM images of PtTe$_2$ films grown from 20 nm Pt precursor with increasing amounts of Te electrodeposited - 2,500, 5,000 and 10,000 pulses. This results in films with increasingly large grains but lower surface coverage of PtTe$_2$. (e) Optical microscopy image of a PtTe$_2$ film with low surface coverage (f) Raman map of the intensity of the E$_g$ peak over the area shown in (e) highlighting that the PtTe$_2$ is present only in certain areas.

Figure 3(e) shows an optical microscopy image of a PtTe$_2$ film (synthesized from 20 nm Pt and 20,000 pulses Te) with low surface coverage. A Raman spectroscopy map of the intensity of the E$_g$ PtTe$_2$ peak over the same area is shown in Figure 3(f). The Raman intensity matches closely the visible crystals, confirming that the PtTe$_2$ is only present in these areas. XPS analysis of the same area, shown in Figure S7 in the SI, indicates that there is no elemental Pt or Te present on the surface of this sample. This implies that during the synthesis both the Pt and Te dewet from the surface leaving only PtTe$_2$ crystals and bare PyC on the surface. The dewetting allows much larger and thicker crystals of PtTe$_2$ to grow, at the expense of complete surface coverage, and is likely promoted by the strong interlayer interactions of PtTe$_2$ and the weak adhesion of Pt to the substrate. This demonstrates that changing the initial precursor quantities offers a way to influence the film thickness, crystal grain size and also surface coverage on the sample. Investigations on the use of different substrates, with different surface energies, to determine their effect on the surface coverage would be a worthwhile future study.

Electrocatalytic Applications

There has been a large volume of promising work investigating the electrocatalytic properties of TMDs for various reactions, such as water splitting.[41-42] The broad trends found for TMD electrocatalysts are that metallic/semimetallic phases tend to be more active, such as 1T versus 2H MoTe$_2$.[43-44] While the basal plane of semiconducting phases is generally reported as being relatively inactive, the creation of defects or functional groups can improve the catalytic behaviour.[45-46] In a similar vein, edge-sites and grain boundaries have been shown to be efficient catalytic sites in many cases, making polycrystalline films of TMDs interesting for electrocatalytic applications.[47] The PtTe$_2$ films grown here have high roughness, conductivity and density of edge-sites making them interesting to examine for electrocatalytic applications.

There has been little work on the electrocatalytic properties of PtTe$_2$ published thus far. The few published studies have demonstrated efficient catalysis of the HER and ORR.[24-26] Although certainly interesting to study, it is important to bear in mind, when touting PtTe$_2$ for catalytic applications, that Pt is an excellent catalyst of numerous reactions in its own right. It is also an expensive and scarce element and, as such, any application of $PtTe_2$ as a catalyst would need to justify the cost of both the raw Pt and the processing. It should be noted however, that $PtTe_2$ is only ~43% Pt by weight, meaning any results that are close to matching Pt would be potentially economically viable if it could be grown efficiently on a large scale.

The HER is an important industrial reaction, which is also well understood, making it an ideal candidate to study the electrocatalytic properties of $PtTe_2$. As previously discussed, some of the films of $PtTe_2$ grown in this study tended to dewet from the substrate and not provide full surface coverage. For this reason, there was a limited range of parameters under which the films could be grown and tested for catalysis of the HER. Only films grown from 10 or 20 nm Pt with 2,500 or 5,000 pulses of Te were examined in this section.

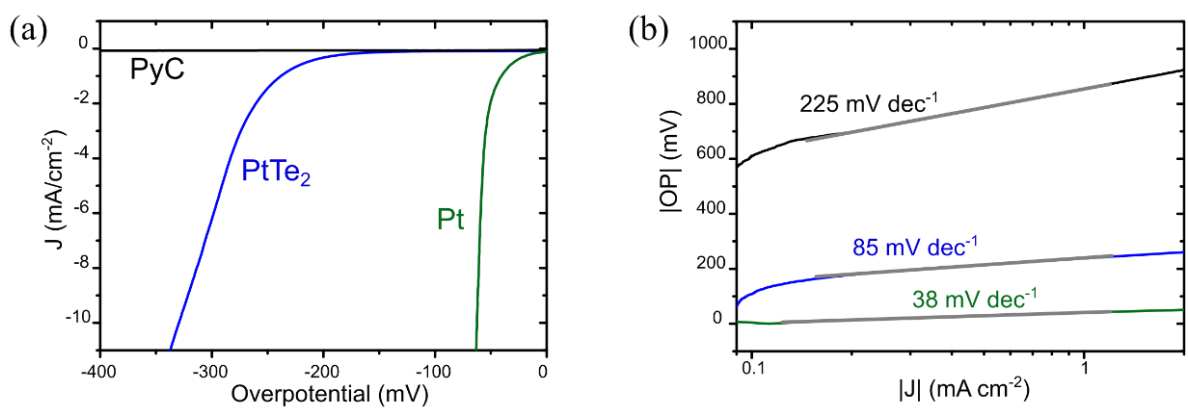

Figure 4 (a) Catalytic activity of the film towards the HER. LSV of a 10 nm $PtTe_2$ film, a Pt only and a PyC only sample. (b) Tafel plot of the same samples with fitted Tafel slopes.

Shown in Figure 4(a) are the linear sweep voltammetry (LSV) results for a 10 nm $PtTe_2$ film, a 30 nm Pt film and for bare PyC. The $PtTe_2$ film demonstrates very strong electrocatalytic behavior with measureable catalytic current at over potentials (OP) as low as -219 mV. A common metric examined in literature to compare the behavior of electrocatalysts of the HER is the OP @ 10 mA $cm^{-2}$. Using this metric the $PtTe_2$ electrodes measured in this work were found to perform well, requiring an OP of ~ - 330 mV to measure a current density of 10 mA $cm^{-2}$. This demonstrates the high activity of the PtTe$_2$ films. The bare PyC shows essentially no activity in this voltage window, while the Pt control shows high catalytic activity as expected.

The Tafel plot for each of these films is shown in Figure 4 (b), this allowed the Tafel slope for the films to be calculated. The PtTe$_2$ film also demonstrates strong behavior by this metric, with a slope of 84 mV dec$^{-1}$. The Pt film shows the expected very low slope of 38 mV dec$^{-1}$, while the PyC only sample shows little activity with a high Tafel slope. XPS analysis of the PtTe$_2$ film after HER measurements is shown in Figure S8 in the SI. There were only minor changes in the composition of the film compared to the pre-HER measurements indicating the stability of the films under these measurement conditions.

Chia *et al.* examined the suitability of PtS$_2$, PtSe$_2$ and PtTe$_2$ as catalysts for the HER.[24] In that work CVT crystals of PtTe$_2$ were synthesized, then exfoliated and subsequently drop-cast onto a glassy carbon electrode for electrochemical measurements. Of the three platinum dichalcogenides studied, PtTe$_2$ was found to be the most active as an electrocatalyst for the HER. They reported an OP of -530 mV @ 10mA cm$^{-2}$ for as synthesized PtTe$_2$. The PtTe$_2$ films examined in the current work required a much lower OP than the material used by Chia *et al*. to achieve the same current density, highlighting their greater electrocatalytic behavior. This improved behavior may be due to the polycrystalline nature of the films, as edge sites and defects have previously been shown to offer improved catalytic performance for Pt based TMDs.[28, 48] The films studied in this work have the added advantage of being directly synthesized on the current collector, a more convenient synthesis process than CVT crystal growth, flake exfoliation and drop-casting. This method also involved a much lower synthesis temperature than typical CVT.

Following the promising HER results, the catalytic activities of the PtTe$_2$ films towards the ORR were investigated. Similar to the HER, the ideal electrocatalyst for the ORR enables the reaction to occur at an appreciable rate as close as possible to the thermodynamically predicted voltage, 1.23 V vs RHE for ORR. Previous reports have highlighted that $PtTe_2$ shows strong catalytic

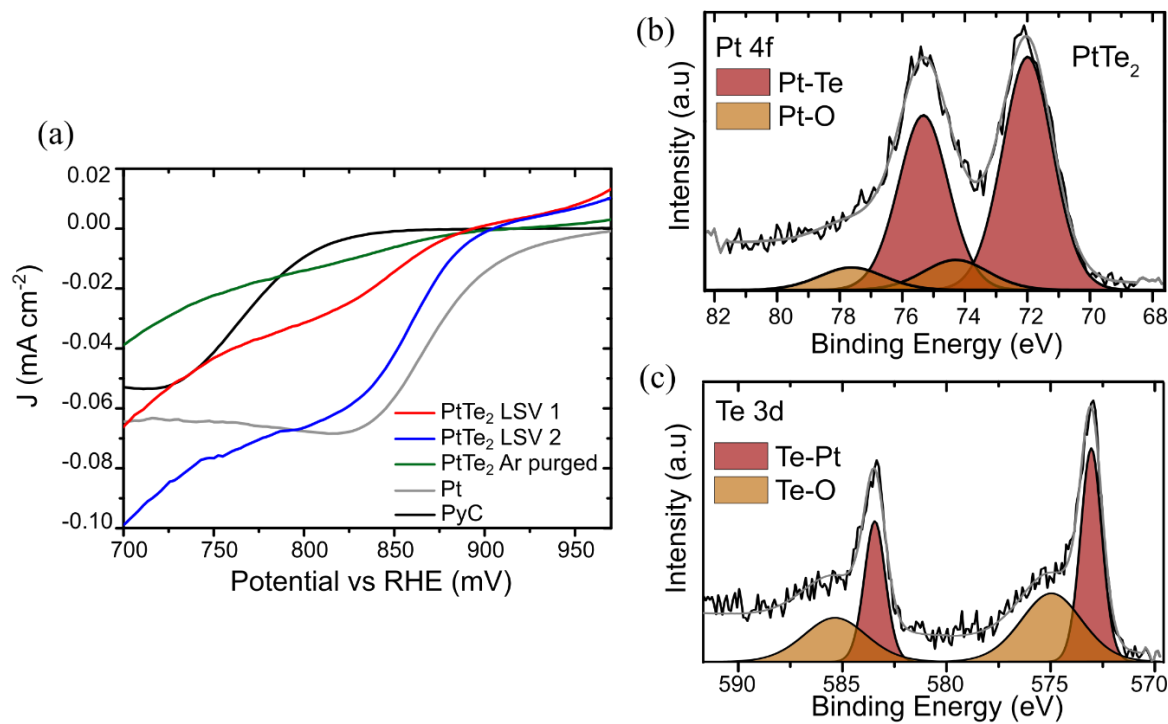

Figure 5 (a) LSV of samples of $PtTe_2$, Pt and PyC with and without Ar purge of electrolyte for the ORR (b) and (c) XPS showing the Pt 4f and Te 3d core level spectra respectively for the $PtTe_2$ sample after the ORR measurements.

behavior, very close to Pt/C, for this system.[26] These previous works were again based on CVT-grown crystals with flakes obtained by liquid-phase exfoliation and drop-cast onto an electrode.

The catalytic activity of a 20 nm $PtTe_2$ film towards the ORR is shown in Figure 5(a). LSV 1 and LSV 2 are two subsequent LSVs of the same $PtTe_2$ film, with LSV 2 performed immediately after LSV 1. The response of the system after the electrolyte was purged with Ar for 20 minutes is also shown. The absence of any peak in this case confirms that the measured values for the other curves are due to the ORR. The response of a Pt film is also shown, along with that of a PyC only film.

There is a dramatic improvement in catalytic activity of the PtTe$_2$ film in LSV 2, while in LSV 1 it did not exhibit strong catalysis of the ORR. Examining LSV 2, its half-wave potential, E$_{1/2}$, was found to be 862 mV, close to the measured E$_{1/2}$ of the Pt film of 876 mV. This demonstrates the similarity in the electrocatalytic activity of both and matches well to the work of Rosli *et al.* where they measured the PtTe$_2$ to be less active than Pt/C by only 18 mV.[26]

The dramatic improvement in the activity of the PtTe$_2$ electrode between LSV 1 and LSV 2 is worth investigating further before drawing firm conclusions. Similar to after the HER measurements, XPS was used to investigate the sample after the ORR measurements. This is shown in Figure 5(b) & (c). It is clear that there are only two states of Pt present in the XPS spectrum - Pt bound to Te and Pt bound to oxygen. This is similar to the spectrum of the as grown PtTe$_2$ films seen in Figure 1(c) & (d), however a much higher level of oxidation is present in this case. It was found that after the ORR almost 50% of the Te on the surface was in the form of an oxide, while for the Pt about 14% of the atoms were bound to oxygen. This increase in oxide is notable, but it is the significant change in the stoichiometry of the PtTe$_2$ that is most surprising. The stoichiometry of the 'PtTe$_2$' on the surface is in fact PtTe$_{0.5}$. This is a huge change from before the ORR measurements and indicates that the sample is now very Pt rich on the surface, to the extent that it could not be described as PtTe$_2$. Despite this, there was no indication of Pt in an elemental state found on the surface within the limits of detection. The lack of elemental Pt signal may imply a single sub-stoichiometric PtTe$_x$ is forming on the surface during the electrochemical measurements and that this material is much more electrocatalytically active than as-synthesized PtTe$_2$. PtO$_x$ has also exhibited catalysis of the HER and so the high level of oxidation must also be considered.[49-50] Previously published work by Rosli *et al.*, in which they study PtTe$_2$ as an electrocatalyst for the ORR, does not mention observing similar Te loss during measurements.[26]

As such, while the PtTe$_2$ films appear relatively stable under the acidic measurement conditions used for HER, which is main proposed electrochemical application of TMDs. However, the ORR follows the trend which was previously observed for Group 6 sulfides and selenidines (MoS$_2$, MoSe$_2$, WS$_2$ and WSe$_2$) which shows that after reductive treatment, there is significant decrease in chalcogen content.[51]

CONCLUSIONS

In this report a low-temperature, straightforward and scalable method for the synthesis of PtTe$_2$ thin films was outlined. The synthesis process involved predeposition of Pt and Te layers that were then reacted under a N$_2$ atmosphere. The synthesis of PtTe$_2$ was confirmed with a number of spectroscopic techniques and microscopy was used to analyze the morphology of the polycrystalline films. The grains of the PtTe$_2$ films were found to be quite 3D potentially due to the strong interlayer bonding in PtTe$_2$ encouraging thicker flake growth.

The synthesis was successful at temperatures between 450°C and 650°C, the lowest growth temperature for PtTe$_2$ currently reported, with all other reported methods requiring temperatures above 700 °C. This opens up the possibility of growth on a wider variety of substrates and potential integration with industry processes.

Finally, the PtTe$_2$ films were examined as electrocatalysts of the HER and the ORR. The films showed strong catalytic behavior for both reactions and compared very favorably to previous studies on CVT-grown crystals of PtTe$_2$. While the films were stable under the HER measurement conditions, they were found to be much more catalytically active towards the ORR after an initial cycling was completed. This change was investigated with XPS and it was found that while there was no evidence of elemental Pt on the surface, the sample surface was very Te deficient. This requires further studies to fully elucidate the behavior.

EXPERIMENTAL METHODS

Synthesis of PtTe$_2$

Films of Pt were deposited onto substrates using a Temescal FC2000 electron-beam evaporation system. This work examined PtTe$_2$ films grown from initial Pt layers between 2 and 40 nm. All samples are named by referring to the thickness of the starting Pt film.

In order to electrodeposit Te onto them, the Pt thin-film samples were used as the working electrode in an electrochemical cell. The Te was reduced from a solution of 0.02 M TeO$_2$ in 1 M nitric acid. A platinum counter electrode and a Ag/AgCl reference electrode were used. The deposition was carried out using a pulsed voltage sequence consisting of 10 ms pulses of -0.2 V applied to the working electrode with respect to the reference electrode with 50 ms gaps between these where no voltage was applied. The reduction proceeded *via* the reaction[52]:

$$HTeO_2^+ + 3H^+ + 4e^- \rightarrow Te + 2H_2O$$

Varying the number of pulses in a deposition sequence served to control the quantity of Te deposited on the sample. Pulsing of the potential was found to result in a more uniform film of Te on the sample.[52] Further details are given in a previous work using a similar method to synthesize MoTe$_2$.[34]

The substrates used for the PtTe$_2$ were pyrolytic carbon (PyC) on a 300 nm thermal SiO$_2$ on Si wafer. The PyC layer served to improve adhesion between the Pt film and the SiO$_2$/Si wafer during the electrodeposition. It also acted as the current collector for electrochemical measurements. The PyC was grown by CVD of acetylene at 950°C for 30 mins on SiO$_2$/Si substrates in a hot-wall, quartz-tube furnace, full details can be found in a previous report.[53]

Following the Te deposition, the samples were annealed in an ATV PEO 604 quartz furnace at a set temperature (450°C unless otherwise specified), allowing the Pt and Te layers react to form PtTe$_2$. The reaction took place under nitrogen atmosphere at a pressure of ~700 mbar. The samples were placed inside nested crucibles in the furnace in order to have a high partial pressure of Te in the vicinity of the samples and to avoid contamination of the furnace. The temperature was ramped at 180 °C min$^{-1}$ and held at the growth temperature for 90 mins. The samples were then allowed to cool to near room temperature (<30 °C) under N$_2$ over a period of ~3 hours before removal from the furnace. During the cooling, at ~320 °C, the pressure was lowered from ~700 mbar to 13 mbar. If the pressure was not lowered, there was found to be an increased likelihood of Te crystals remaining on the surface of the film.

Sample Characterization

A WITec Alpha 300R with a 532 nm excitation laser, with a power of ~200 µW, was used to collect the Raman spectra. All Raman measurements were taken using a spectral grating with 1800 lines/mm and a 100x objective lens (N.A. = 0.95). All spectra shown are averages of at least 20 points, each with an integration time of 3 seconds.

X-ray photoelectron spectroscopy (XPS) spectra in Figure 1, S7 and S8 were taken using a PHI VersaProbe III instrument equipped with a micro-focused, monochromatic Al Kα source (1486.6 eV) and dual beam charge neutralization. Core-level spectra were recorded with a spot size of 100 µm and a pass energy of 69 eV using PHI SmartSoft VersaProbe software, and processed with PHI MultiPak. Binding energies were referenced to the adventitious carbon signal at 284.8 eV.

After subtraction of a Shirley type background, the spectra were fitted with Gaussian–Lorentzian peak shapes.

The XPS in Figure 5 was carried out using a monochromated ESCAProbeP spectrometer (Omicron Nanotechnology Ltd, Germany) with an aluminium X-ray radiation source (1486.7 eV). The survey scans were carried out using a pass energy of 50 eV. The high resolution core level scans were conducted using a pass energy of 30 eV. All analysis was performed using CasaXPS software.

Scanning electron microscopy (SEM) images were obtained with a Karl Zeiss Supra microscope operating at 3 kV accelerating voltage, 30 μm aperture and a working distance of ~3-4 mm.

Atomic force microscopy (AFM) was carried out on a Bruker Multimode 8 in ScanAsyst Air mode using Nanosensor PointProbe Plus tips.

The X-ray diffraction (XRD) measurement was performed on a Bruker D8 Discover with a monochromated Cu K-alpha source.

STM measurements were taken on a commercial low-temperature STM from Createc Company with a base pressure of $3 \times 10^{-11}$ mbar. All STM images were recorded at liquid nitrogen temperature (77 K) in constant-current mode. The bias was applied to the sample with respect to the tip. The STM tip was (001)-oriented single-crystalline tungsten, which was electrochemically etched in NaOH

Electrochemical Measurements

The samples were measured for the HER in a three-electrode electrochemical cell with sulfuric acid (0.5 M) as the electrolyte with the PtTe$_2$ film as the working electrode, a large graphitic counter electrode and a Ag/AgCl reference electrode. Catalytic activity was measured by performing linear sweep voltammetry (LSV) and electrochemical impedance spectroscopy (EIS) with either a Gamry Reference 3000 or 600 potentiostat at a potential of 0 mV vs RHE.. Linear voltage sweeps were performed at a scan rate of 5 mV s$^{-1}$ in a voltage range -0.2 V to −1.2 V (vs Ag/AgCl). EIS in the frequency range of 0.1 to 10 MHz with perturbation voltage amplitude of 10 mV was used to determine the equivalent series resistance of the system. All HER data was corrected for the electrolyte resistance by iR compensation.

The ORR measurements were carried out using an Autolab potentiostat (series no. PGSTAT204). ORR measurements were performed in a three-electrode electrochemical cell with 1M NaOH as the electrolyte, a graphitic counter electrode and a Hg/HgO reference electrode. For the 'Ar purged PtTe$_2$' sample, the electrolyte was purged with argon gas for 20 minutes prior to ORR testing.

ASSOCIATED CONTENT

**Supporting Information**.

Diagram of electrodeposition; Fitted PtTe$_2$ Raman spectrum; Table of published PtTe$_2$ Raman spectroscopy FWHM values; AFM images of PtTe$_2$ films; Raman and XPS measurements of samples grown at high temperature; Pt-Te phase diagram (adapted from literature); Plots of film morphology variation with Pt and Te quantities; XPS of PtTe$_2$ film with low surface coverage; XPS of film post HER measurement;

AUTHOR INFORMATION

**Corresponding Author**

*E-mail nmcevoy@tcd.ie

**Author Contributions**

The manuscript was written through contributions of all authors. All authors have given approval to the final version of the manuscript.

ACKNOWLEDGMENT

J.B. Mc M. acknowledges an Irish Research Council scholarship, Project 204486, Award 13653. N. M. acknowledges support from SFI through 15/SIRG/3329. G. S. D. acknowledges the support of SFI under Contract No. 12/RC/2278 and PI_15/IA/3131.  M.P.B. would like to acknowledge the European Structural and Investment Funds, OP RDE-funded project "ChemJets" (No. CZ.02.2.69/ 0.0/0.0/16_027/0008351). M.P. acknowledges the financial support of Grant Agency of the Czech Republic (EXPRO: 19-26896X). The SEM imaging for this project was carried out at the Advanced Microscopy Laboratory (AML),Trinity College Dublin, Ireland. The AML (www.tcd.ie/crann/aml) is an SFI supported imaging and analysis centre, part of the CRANN Institute and affiliated to the AMBER centre.

# Low-temperature synthesis and electrocatalytic application of large-area PtTe$_2$ thin films

*John B. Mc Manus*[1,2], *Dominik V. Horvath*[1,2], *Michelle P. Browne*[3], *Conor P. Cullen*[1,2], *Graeme Cunningham*[1,2], *Toby Hallam*[4], *Kuanysh Zhussupbekov*[5], *Daragh Mullarkey*[5], *Cormac Ó Coileáin*[1,2], *Igor V. Shvets*[5], *Martin Pumera*[3,6,7], *Georg S. Duesberg*[1,8], *Niall McEvoy\**[1,2]

[1] School of Chemistry, Trinity College Dublin, Dublin 2, D02 PN40, Ireland

[2] AMBER Centre, CRANN Institute, Trinity College Dublin, Dublin 2, D02 PN40, Ireland

[3] Center for Advanced Functional Nanorobots, Department of Inorganic Chemistry, University of Chemistry and Technology Prague, Technicka 5, 166 28 Prague 6, Czech Republic

[4] Emerging Technologies and Materials Group, School of Engineering, Newcastle University Merz Court, Newcastle Upon Tyne, NE1 7RU. United Kingdom

[5] School of Physics, Trinity College Dublin, Dublin 2, D02 PN40, Ireland

[6] Department of Medical Research, China Medical University Hospital, China Medical University, No. 91 Hsueh-Shih Road, Taichung, Taiwan

[7] Future Energy and Innovation Laboratory, Central European Institute of Technology, Brno University of Technology, Purkyňova 656/123, Brno, CZ-616 00, Czech Republic

[8] Institute of Physics, EIT 2, Faculty of Electrical Engineering and Information Technology, Universität der Bundeswehr, 85579 Neubiberg, Germany

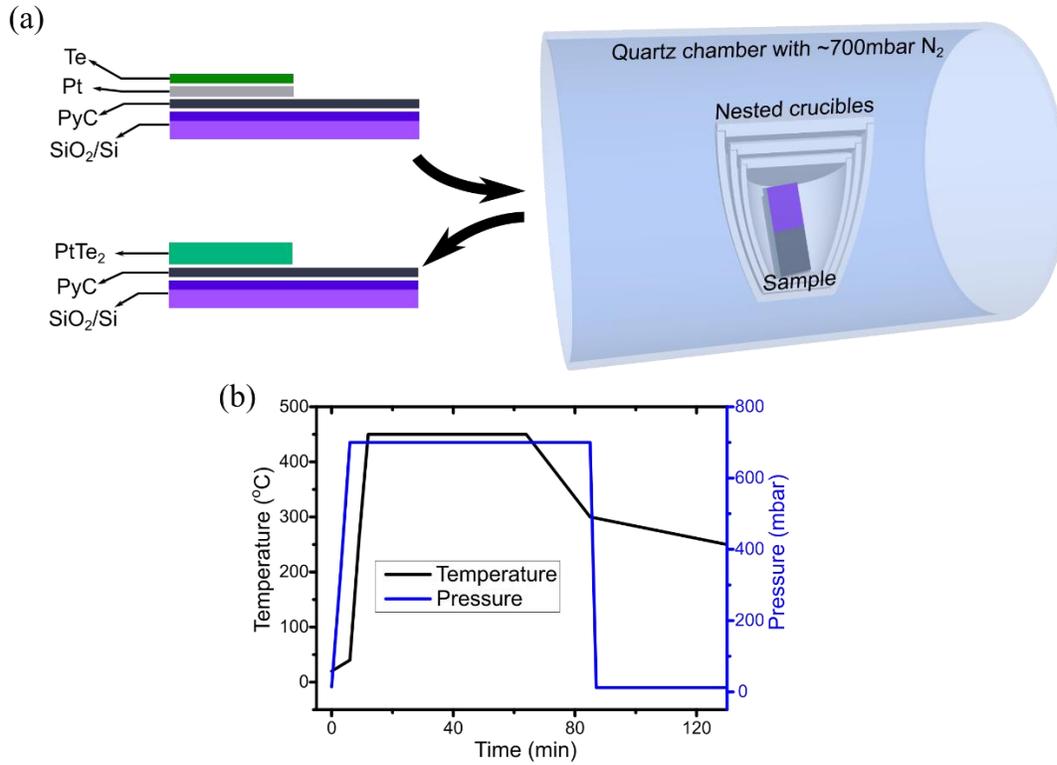

Figure S1. (a) Diagram of the furnace setup showing the sample geometry and the nested crucibles in a quartz tube furnace. The nested crucibles served to keep the local concentration of Te high during the reaction, this was also aided by the 700 mbar of $N_2$ pressure in the furnace during the growth. (b) Graph showing the temperature and pressure inside the furnace during the growth of the $PtTe_2$ from the precursor Pt and Te layers.

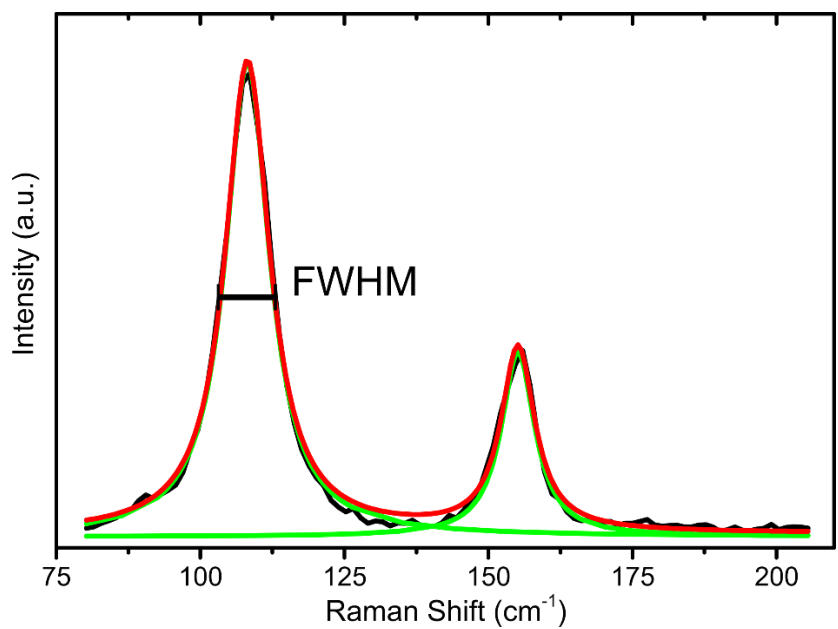

Figure S2: PtTe$_2$ Raman spectrum with the E$_g$ and A$_{1g}$ peaks fitted with Lorentzian peak shapes allowing the full width at half maximum (FWHM) of each to be extracted.

Table S1: Comparison of the FWHM values of the Raman modes of PtTe$_2$ in this work and other previously published literature. Values for other works were estimated from digitized graphs. The values shown would be influenced to some extent by the measurement setup in each case; however, exact details on this were not present in the published reports.

| Reference | FWHM E$_g$ mode (cm$^{-1}$) | FWHM A$_{1g}$ mode (cm$^{-1}$) |
|---|---|---|
| Hao et al.[1] | 9.2 | 6.6 |
| Yan et al.[2] | 9.5 | 7.9 |
| Fu et al.[3] | 10.2 | 7.7 |
| This work | 6.9 ± 0.5 | 4.7 ± 0.4 |

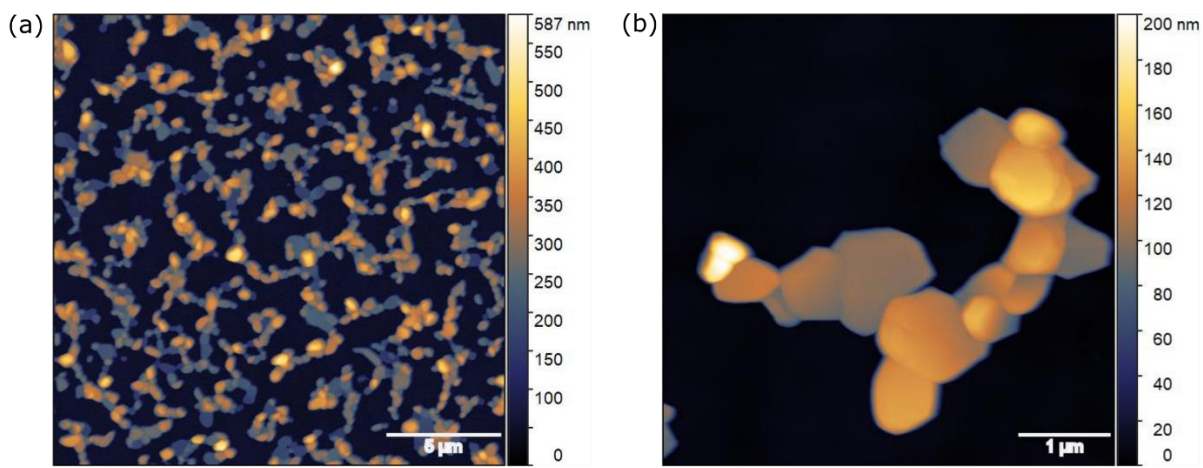

Figure S3: AFM images showing the 3D nature of the individual PtTe$_2$ flakes. This analysis was carried out on quite thick samples (precursor layers: 20 nm Pt, 10,000 pulses Te) as they tended to dewet from the substrate and so facilitate accurate height measurements of the flakes.

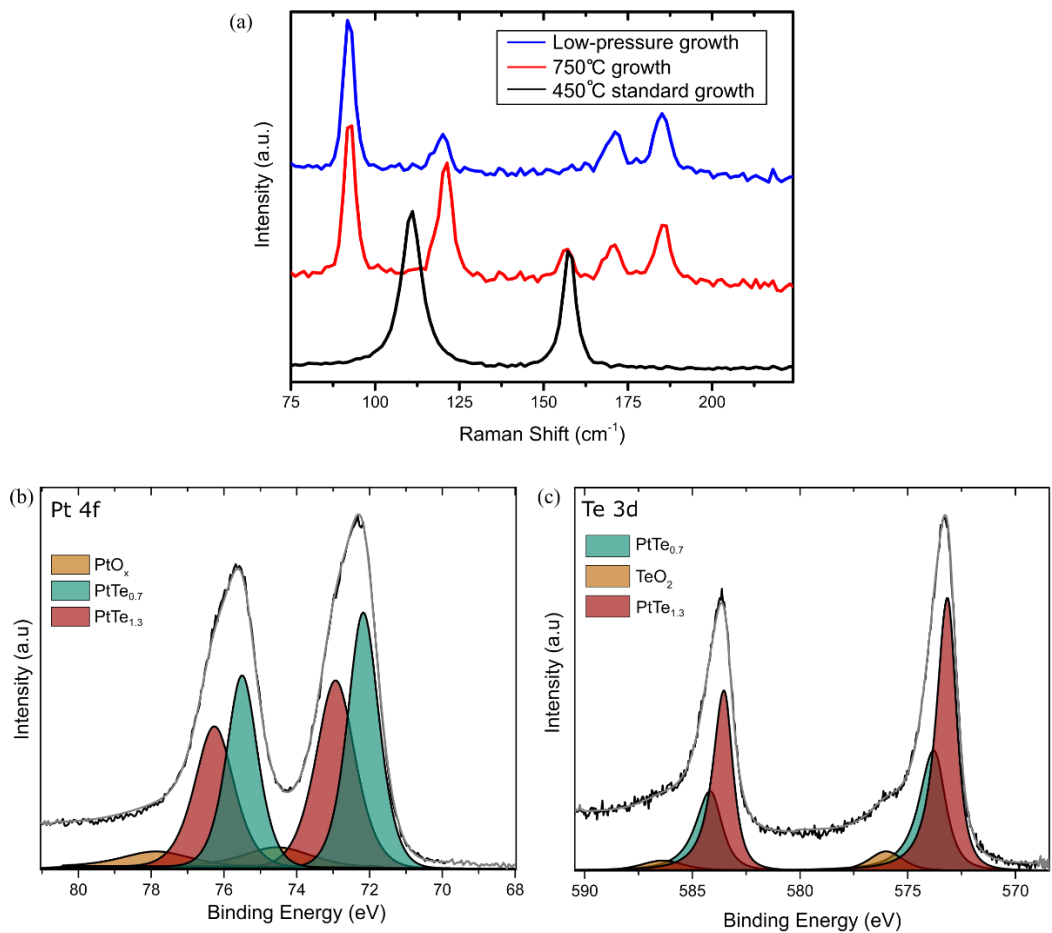

Figure S4: Raman spectra of samples from three different growths, one at 450 °C under standard conditions (700 mbar $N_2$, 90 min growth time) and one at 750 °C with otherwise standard growth conditions. The third growth was carried out at 450 °C, under 13 mbar of $N_2$ with no Te deposited on the growth substrate. Both the 750 °C and low-pressure growths gave very similar Raman spectra with a number of peaks, as yet unidentified in literature. This suggests that at 750 °C a different Pt-Te phase is being formed. To investigate this further, XPS analysis was completed on the 750 °C sample, the Pt 4f and Te 3d region are shown in (b) and (c) respectively. The XPS shows the Pt and Te oxides along with two components representative of Pt bound to Te. The stoichiometry of these are $PtTe_{0.7}$ and $PtTe_{1.3}$. This confirms our expectation of the formation of less Te-rich phases at higher growth temperatures. There are a number of possible phases these could be, as shown by the phase diagram in Figure S4, however no designation could be definitively made based on this analysis.

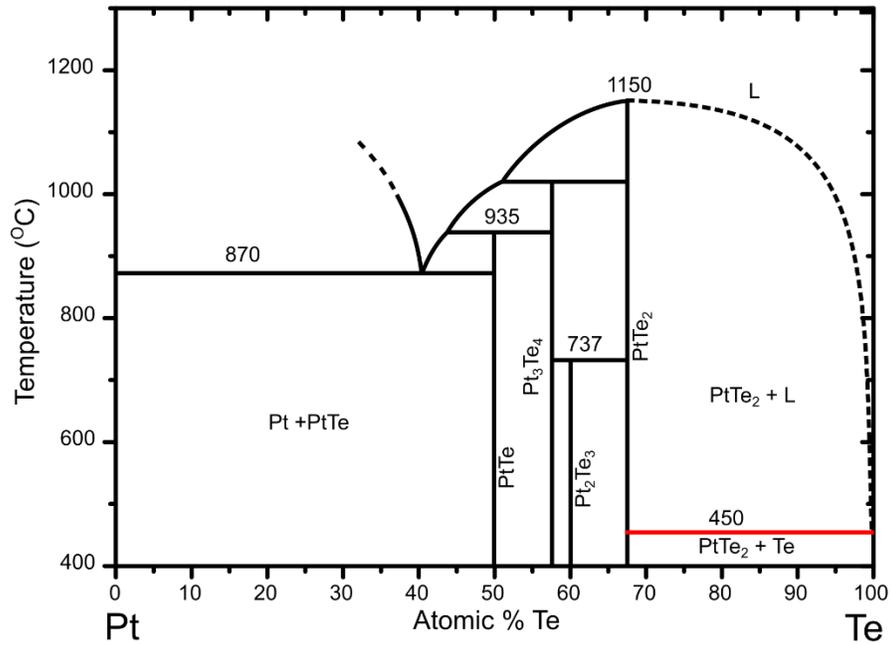

Figure S5: Pt-Te phase diagram reproduced from ref.[4] The line along which the reaction happens in this system is colored red. The phase diagram shows that a Pt-Te solution with low concentrations of Pt can be formed just above 450 ˚C.

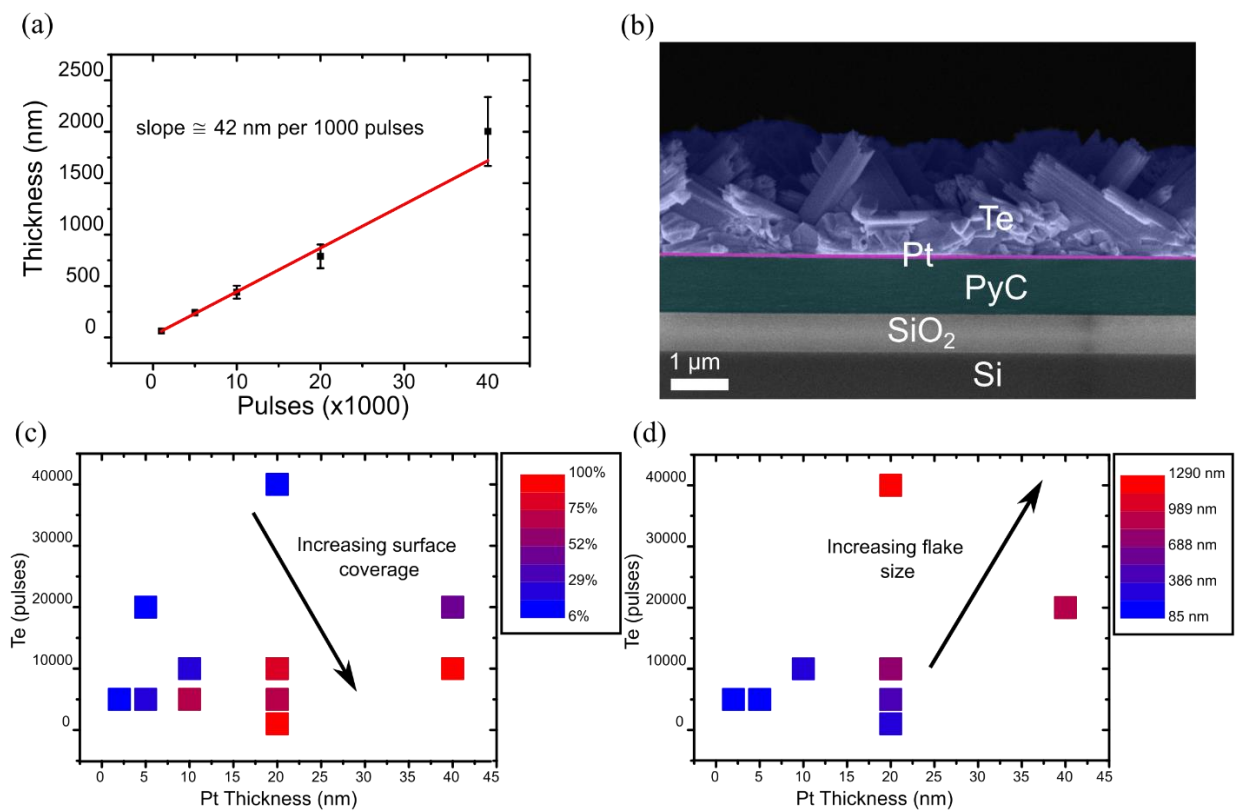

Figure S6: (a) Plot of the relationship between the thickness of the Te film deposited on the surface and the number of Te pulses used during the electrodeposition. The thickness was measured using cross-sectional SEM, an example of which is shown in (b). The SEM also shows that the Te electrodeposition results in long facets or rods of Te on the surface, giving a very porous Te film. This means that the relative thicknesses of the Pt and Te layers do not give a good indication of the relative quantities of the reactants on the surface.

To investigate the variation in film morphology the initial quantity of Pt and Te on the surface of the samples was changed in a systematic fashion, with all other growth parameters kept constant. Shown in (c) is a plot of the results of this experiment. Each point is a sample with a defined initial quantity of Pt and Te on the surface. The color of a point is related to the percentage of the surface that is covered by $PtTe_2$ crystals after the growth. It is clear that increasing the Te amount relative to the Pt thickness reduces the surface coverage. (d) Plot of the effect of the same parameters (quantity of Pt and Te) on the average lateral flake size in the $PtTe_2$ films. The size of the crystals of $PtTe_2$ in the film is similarly shown by the coloring of each data point. The flake size increases as the relative amount of Te increases.

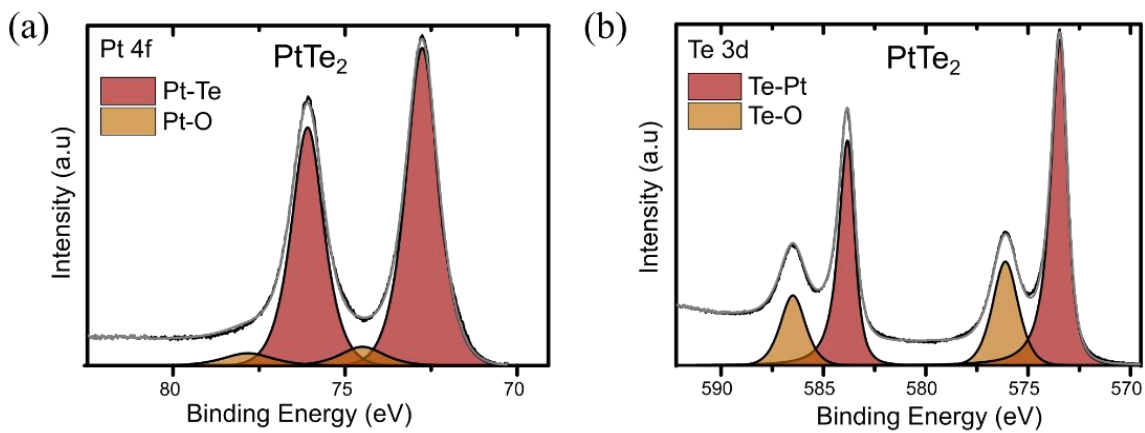

Figure S7 (a) XPS of sample partially covered with PtTe$_2$ crystals showing the two states of Pt present. These correspond to Pt bound to Te and O. There is no evidence of elemental Pt in the sample, which would be expected to be at a lower binding energy than the observed peaks. (b) The equivalent spectra of the Te region.

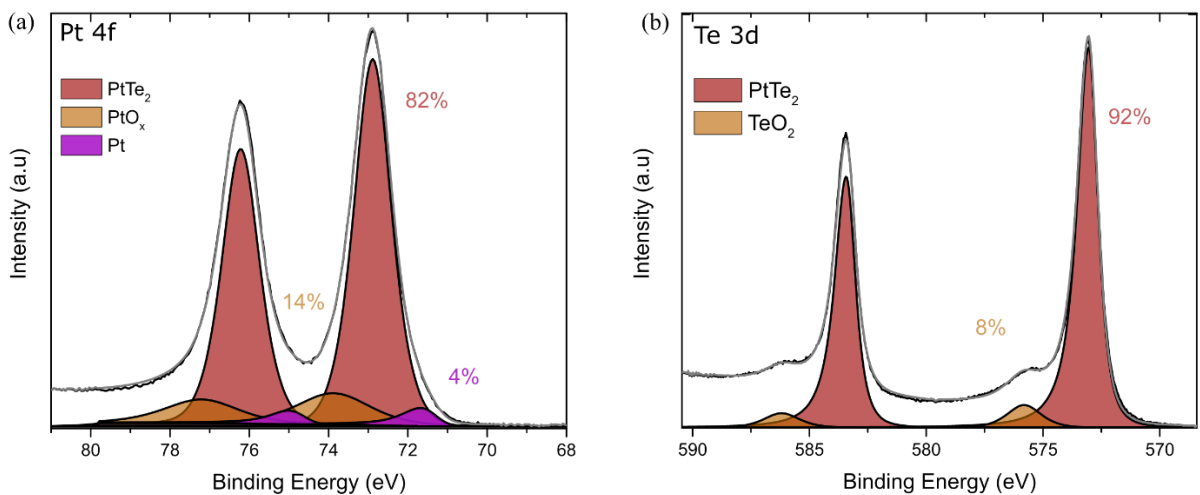

Figure S8 (a) XPS of the Pt 4f region from a PtTe$_2$ sample after it was used for the HER measurements. (b) The associated Te 3d region. These spectra show similar levels of oxidation on the surface as pristine films, about 10%. The stoichiometry is also unchanged at PtTe$_{2.2}$. The only significant difference is the development of a small elemental Pt component (4%) after the HER measurement. This highlights the relative stability of the films under the HER measurement conditions.